\documentclass[final,10pt,times]{article}

\usepackage[T1]{fontenc}
\usepackage[utf8]{inputenc}

\usepackage{cite}
\usepackage{microtype}

\DisableLigatures[f]{encoding = *, family = * }

\raggedright
\usepackage{hhline}
\usepackage{array}
\usepackage{multirow}
\usepackage{booktabs}
\usepackage{tabularx,hhline}
\usepackage{float}
\usepackage{amsmath}
\usepackage{gensymb}
\usepackage{color,soul} 
\usepackage{mathtools, cuted}
\usepackage{lipsum, color}

\usepackage{changepage}
\usepackage[aboveskip=1pt,labelfont=bf,labelsep=period,singlelinecheck=off]{caption}

\makeatletter
\renewcommand{\@biblabel}[1]{\quad#1.}
\makeatother

\usepackage{lastpage,fancyhdr,graphicx}
\usepackage{epstopdf}
\fancyheadoffset[L]{2.25in}
\fancyfootoffset[L]{2.25in}

\usepackage{color}

\definecolor{Gray}{gray}{.25}

\usepackage{graphicx}

\usepackage{sidecap}

\usepackage{wrapfig}
\usepackage[pscoord]{eso-pic}
\usepackage[fulladjust]{marginnote}
\reversemarginpar

\begin{document}
\vspace*{0.35in}

\begin{flushleft}
{\Large
\textbf\newline{SENSE: A comparison of photon detection efficiency and optical crosstalk of various SiPM devices}
}

A. Nagai\textsuperscript{1, *},
C. Alispach\textsuperscript{1},
T. Bergh\"ofer\textsuperscript{2},
G. Bonanno\textsuperscript{3},
V. Coco\textsuperscript{1},
D. della Volpe\textsuperscript{1},
A. Haungs\textsuperscript{4},
M. Heller\textsuperscript{1},
K. Henjes-Kunst\textsuperscript{2},
R. Mirzoyan\textsuperscript{5},
T. Montaruli\textsuperscript{1},
G. Romeo\textsuperscript{3},
Y. Renier\textsuperscript{1},
H.C. Schultz-Coulon\textsuperscript{6},
W. Shen\textsuperscript{6},
D. Strom\textsuperscript{5},
H. Tajima\textsuperscript{7},
I. Troyano-Pujadas\textsuperscript{1},
\\
\bigskip
\bf{1} Department de physique nucléaire et corpusculaire, Universite de Geneve, 24 Quai E. Ansermet, Switzerland
\\
\bf{2} DESY - Deutsches Elektronen-Synchrotron, Notkestrasse 85, 22607, Hamburg, Germany
\\
\bf{3} INAF- Catania Astrophysical Observatory, Via S. Sofia, 78 I-95123 Catania, Italy
\\
\bf{4} KIT - Karlsruhe Institute of Technology, Karlsruhe, 76021, Germany
\\
\bf{5} MIT - Max-Planck-Institute for Physics, Foehringer Ring 6 80805 Munich, Germany
\\
\bf{6} University of Heidelberg, Im Neuenheimer Feld 227 Kirchhoff Institute for Physics 69120 Heidelberg Germany
\\
\bf{7} Nagoya University, Furo-cho, Chikusa-ku, Nagoya 464-8601, Japan

\bigskip
* Andrii.Nagai@unige.ch

\end{flushleft}

\begin{abstract}

This paper describes a comparison of photon detection efficiency and optical crosstalk measurements performed
by three partners: Geneva University, Catania Observatory and Nagoya University. The measurements were compared for three different SiPM devices with different active areas: from 9 $mm^2$ up to 93.6 $mm^2$ produced by Hamamatsu. The objective of this work is to establish the measurements and analysis procedures for calculating the main SiPM parameters and their precision. This work was done in the scope of SENSE project which aims to build roadmap for the last developments in field of sensors for low light level detection.

\end{abstract}


\section{Introduction}
\label{Sec:Introduction}

SENSE \cite{SENSEPage} is a coordinated research and development consortium between academic research groups and industry with the common goal of developing the ultimate low light level (LLL) sensor. It is funded
by the European Commission under Future and Emerging Technologies (FET) Open Coordination and Support Action (CSA). 

The project\textsc{\char13}s objectives are: (1) to conduct the development of a European $R\&D$ roadmap towards the ultimate LLL sensors, and to monitor and evaluate the progress of the development with respect to the roadmap, (2) to coordinate the $R\&D$ efforts of research groups and industries in advancing LLL sensors and liaise with strategically important European initiatives and research groups and companies worldwide, (3) to transfer knowledge by initiating information and training events and material, (4) to disseminate information by suitable outreach activities.

The consortium has four partners: Deutsches Elektronen Synchrotron (Coordinator), Germany; Universite de Geneve, Switzerland; Max-Planck Institute for Physics, Germany and Karlsruhe Institute of Technology, Germany. Several international experts on all parts of LLL developments are involved in the expert or working group of the project.

\section{Cross characterization challenge in the frame of SENSE}
 
A cooperation agreement between the SENSE Consortium (See section \ref{Sec:Introduction}), the Catania Astrophysical Observatory, of INAF, the University of Heidelberg and the Nagoya University is being established. Several international experts, who are involved in key parts of the LLL development, are involved in the project's experts working group. From this agreement, the invited institutes are working to establish the measurements and analysis procedures for the main SiPM parameter measurements and associated precision. Furthermore, the agreement will facilitate collaboration between SiPM producers on new developments and comparing their performances. 

Initially, the partners of the agreement compared the measurement procedures and established the precision of the different experimental set-ups. Therefore, a few benchmarks of SiPM devices were measured:
\begin{itemize}
\item large area hexagonal SiPM with 50 $\times$ 50 $\mu m$ micro-cell size and 93.6 $mm^{2}$ active area, which is being deployed to build gamma-ray cameras suitable for the Cherenkov Telescope Array Observatory \cite{Inoue13}. This device is produced by Hamamatsu in collaboration with the University of Geneva \cite{CameraPaperHeller2017};
\item two Hamamtsu low voltage reverse series devices, with 50 $\times$ 50 $\mu m$ micro-cell size:
	\begin{itemize}
	\item 3 $\times$ 3 $mm^2$ device, LVR-3050CS, S/N2;
	\item 6 $\times$ 6 $mm^2$ device, LVR-6050CS, S/N7;
	\end{itemize}	 
\end{itemize}

Until now, the partners of the agreement concentrated on precision measurements of optical crosstalk $P_{XT}$ and photon detection efficiency $PDE$ as a function of overvoltage and wavelength of SiPM devices at room temperature. The experimental apparatus and results are presented in the following.

\section{Experimental setup at IdeaSquare/UNIGE}

The experimental setup at the IdeaSquare \cite{IdeaSquarePage}, CERN was build and calibrated in collaboration with the University of Geneva to characterize electrical and optical properties of SiPM devices at room temperature. 

This setup allows both static DC and dynamic AC tests of various SiPM detectors under dark or light illumination conditions and at different wavelengths. All measurements are automatized through a LabView framework.
For DC measurements (i.e. reverse and forward IV), the SiPM device is directly connected to a Keithley 6487 picoammeter for bias supply and current measurements. A 75 W Xe lamp coupled with a monochromator (ORIEL Instruments TLS-75X) was used as a variable wavelength light source (from 260 nm to 1200 nm). 

The data acquisition system for the AC measurements was designed around a pre-amplifier based on the operational amplifier OPA846, a Lecroy 610Zi oscilloscope to acquire the waveform, and a Keithley 6487
picoammeter to supply bias voltage to the SiMP. 

As a source of pulsed light, the LED biased by pulse generator can be used. The LED of various wavelengths are available: 280, 340, 375, 405, 420, 455, 470, 505, 525, 530, 565, 572 and 630 nm. 
A calibrated Hamamatsu S1337-1010BQ photodiode is used to measure the light intensity. The incoming light is spread between the SiPM and the photodiode by an integrated sphere (Thorlab, Model IS200-4). To reduce the amount of light reaching the SiPM, an absorptive Neutral Density Filter (Thorlab, Model NE530B) ND Filter is mounted between the integrating sphere`s output port and the SiPM. A 50$ \degree$ Square Engineered Diffuser (Thorlab, ED1-S50-MD) was mounted after the ND Filter to uniformly illuminate the full active area of the SiPM.



\section{Measurements}

\subsection{Photon detection efficency $PDE$ vs. Overvoltage $\Delta V$}
\label{Sec:PDEAC}

\begin{figure}
\begin{center}\includegraphics[%
  width=8.2cm,
  keepaspectratio]{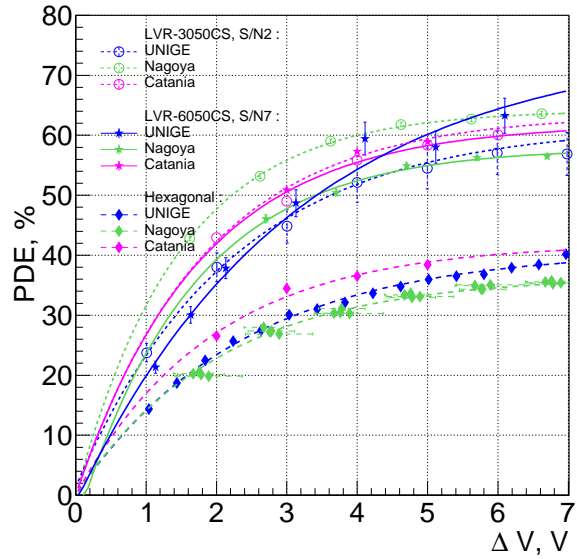}\end{center}
\caption{$PDE$ as a function $\Delta V$ at 405 nm wavelength for three Hamamatsu devices, measured by $Catania$, $Nagoya$ and $UNIGE$ at room temperature.}
\label{Fig:PDEAll}
\end{figure}

All partners agreed to compare the photon detection efficency $PDE$ at a common wavelength $\lambda$ = 405 nm. The absolute $PDE$ as a function of overvoltage $\Delta V$ was calculated using a so-called Poisson method \cite{ECKERT2010217} with a correction for the uncorrelated noise applied. The experimental data were parametrized as:

\begin{multline}
 \label{Eq:PDE}
 PDE(\Delta V, \lambda) = QE (\lambda) \times \epsilon \times P_{Geiger}(\Delta V) = \\
 = PDE_{max}(\lambda) \times \left( 1 - exp \left( PDE_{slop} (\lambda) \cdot \Delta V  \right)  \right)
\end{multline}

where $QE(\lambda)$ is the quantum efficency, $\epsilon$ is the geometrical fill factor, $P_{Geiger}(\Delta V)$ is triggering probability, $PDE_{max}(\lambda) = QE(\lambda) \times \epsilon$, and $PDE_{slop}(\lambda)$ is the parameter depending on the SiPM design and composition of free carriers. Results are presented in Fig. \ref{Fig:PDEAll}, a parameterisation provides a good description of experimental data for all measured devices. This parametrization was used to calculate the relative difference between $PDE$ values calculated by the three partners as:

\begin{equation}
 \label{Eq:PDEResiduals}
 100 \% \times  \frac{\Delta PDE(\Delta V)}{PDE(\Delta V)} \underset{ i \neq j} =  100 \% \times  \frac{PDE_{i}(\Delta V) - PDE_{j}(\Delta V)}{PDE_{i}(\Delta V)}
\end{equation} 

where $i$ and $j$ are partners names = $\left[ UNIGE, \ Catania, \ Nagoya \right]$, $PDE_{i}(\Delta V)$ is the fit function for a data from a given partner at a given $\Delta V$. The $100 \% \times  \frac{\Delta PDE(\Delta V)}{PDE(\Delta V)}$ was calculated at $\Delta V$ range from 0.5 V up to 6 V and we found that on average $100 \% \times  \frac{\Delta PDE(\Delta V)}{PDE(\Delta V)}$ = 7.8 $\%$ relative difference.


\subsection{Photon detection efficency PDE vs. wavelength $\lambda$}

The $PDE$ at various wavelengths was measured and compared by $UNIGE$ and $Catania$ for LVR-3050CS and Hexagonal devices.

\begin{figure}
\begin{center}\includegraphics[%
  width=8.2cm,
  keepaspectratio]{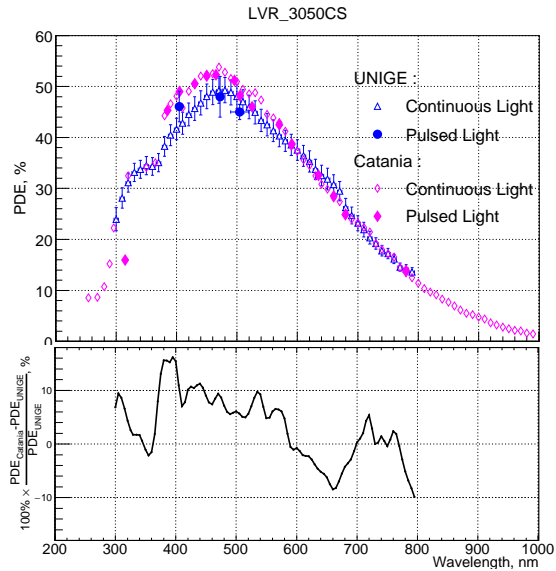}\end{center}
\caption{Cross-check of PDE as a function of wavelength at 3 V overvoltage. Results were obtained by University of Geneva and Catania Observatory. On average the relative difference of 6.4 $\%$ was found.}
\label{Fig:PDEvsWavelength}
\end{figure}

The relative shapes of $PDE$ vs. $\lambda$ were normalized to absolute $PDE$ measured with pulsed light (see previous subsection \ref{Sec:PDEAC}) and presented in Fig. \ref{Fig:PDEvsWavelength}. In average the relative difference of 6.4 $\%$ was found.

\begin{figure}
\begin{center}\includegraphics[%
  width=8.2cm,
  keepaspectratio]{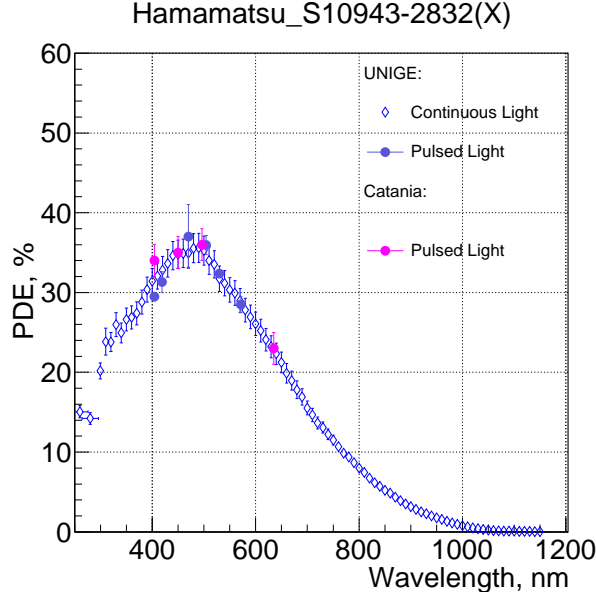}\end{center}
\caption{Cross-check of PDE as a function of wavelength at 3 V overvoltage. Results were obtained by University of Geneva and Catania Observatory. In average the relative difference of 1.8 $\%$ was found for four wavelengthes: 405, 450, 496, 635 nm.}
\label{Fig:PDEvsWavelengthHex}
\end{figure}

The $PDE$ for Hexagonal device was compared at four wavelengthes ($\lambda$ = 405, 450, 496 and 635 nm) at 3 V overvoltage. The relative shape of $PDE$ vs. $\lambda$ was normalized to absolute $PDE$ measured with pulsed light by $UNIGE$ while only absolute $PDE$ at four wavlengthes was measured by $Catania$ (See Fig. \ref{Fig:PDEvsWavelengthHex}). The average relative difference of $PDE$ for those four wavelengths was found to be 1.8 $\%$.

\subsection{Optical crosstalk $P_{XT}$ vs. $\Delta V$}

Assuming cross-talk probability $P_{XT}$ is responsible for a dark count rate at the level of 1.5 photoelectrons $p.e.$ threshold, it can be calculated as:

\begin{equation}
 \label{Eq:PXT}
 P_{XT} =  \frac{ DCR_{1.5 p.e.} } { DCR_{0.5 p.e.} }
\end{equation}
where $DCR_{0.5 p.e.}$ and $DCR_{1.5 p.e.}$ are the dark count rates at 0.5 $p.e.$ and 1.5 $p.e.$ threshold respectively. $UNIGE$ correct the $P_{XT}$ for pile-up effects (caused by thermally generated carriers) as:

\begin{equation}
 \label{Eq:PXTCorr}
 P_{XT} =  \frac{ DCR_{1.5 p.e.} - 2 \cdot \tau \cdot DCR_{0.5 p.e.}^{2} } { DCR_{0.5 p.e.} + 2 \cdot \tau \cdot DCR_{0.5 p.e.}^{2}}
\end{equation}
where $\tau$ is the minimum time interval between two SiPM pulses within which the pulses can be recognised as separated (for $UNIGE$ the $\tau$ = 2 ns.).
The $Nagoya$ group calculated the $P_{XT}$ from Poisson statistics, with the correction for pile-up effect, as:

 \begin{equation}
 \label{Eq:PXTNagoya}
 P_{XT} =  \frac{N_{>1.5 p.e.}}{\mu P(0) \cdot N_{total}} - \frac{\mu}{2} - \frac{\mu^{2}}{6}
\end{equation}

where $N_{total}$ and $N_{>1.5 p.e.}$ are the total number of events and number of events which crosses a threshold of 1.5 $p.e.$ respectively, $\mu$ - is an average from Poisson distribution, $P(0)$ - is probability to have 0 p.e., $ \frac{\mu}{2}$ and $ \frac{\mu^{2}}{6}$ are the probabilities to have pile-up effects from two and three thermal pulses, respectively.

Optical cross-talk occurs when external photons are emitted during the primary avalanche multiplication process. This is due to hot carrier luminescence \cite{PxtMeasurements} and starts secondary avalanches in one or more neigboring micro-cells. Therefore, the $P_{XT}$ was approximated as:

\begin{equation}
 \label{Eq:PXTFit}
 P_{XT} =  \frac{ C_{\mu cell} \cdot \Delta V } { e } \times P_{h \nu} \times P_{Geiger}
\end{equation}

where $P_{Geiger}$ is the Geiger triggering probability, $P_{h \nu}$ is the probability that external photons will be emitted, $ \frac{ C_{\mu cell} \cdot \Delta V } { e }$ is the number of charges created during primary avalanche multiplication ($C_{mu cell}$ - is the SiPM micro-cell capacitance, $ \Delta V$ - is the overvoltage).

\begin{figure}
\begin{center}\includegraphics[%
  width=8.2cm,
  keepaspectratio]{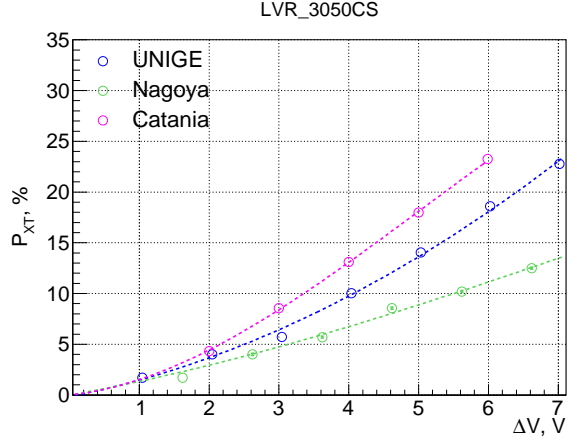}\end{center}
\caption{First results of cross-check of $P_{XT}$ as a function of Overvoltage, between three partners at room temperature.}
\label{Fig:PxtFirstResults}
\end{figure}

First results show large differences (up to 100$\%$) between $P_{XT}$ calculated by SENSE team (an example for LVR-3050CS device is presented in Fig. \ref{Fig:PxtFirstResults}). This difference led to the improvement on measurement setups and data analysis. The most correct value of $P_{XT}$ is presented by $Nagoya$. The results from two other partners were overestimated due to pile-up effects. To reduce a pile-up effect new measurements at much higher signal processing bandwidth were performed (1 GHz instead of 20 MHz) as well as an off-line correction procedure was applied. 

\begin{figure}
\begin{center}\includegraphics[%
  width=8.2cm,
  keepaspectratio]{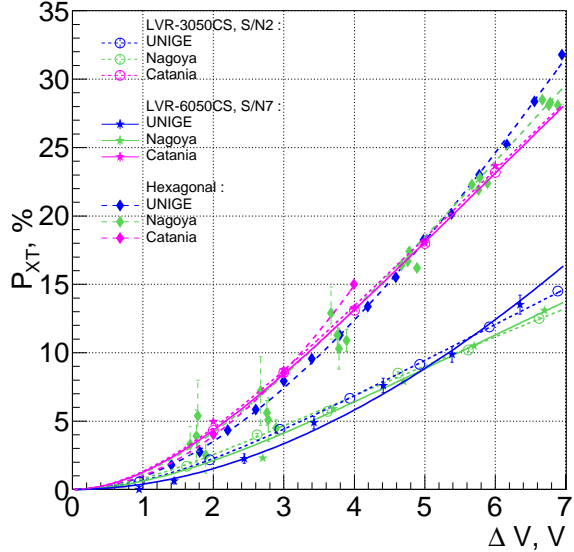}\end{center}
\caption{$P_{XT}$ as a function $\Delta V$ for three Hamamatsu devices, measured by three partners of Agreement at room temperature.}
\label{Fig:PxtAll}
\end{figure}

The new results are presented in Fig. \ref{Fig:PxtAll}. We can observe a good agreement between $P_{XT}$ calculated by all partners for Hexagonal device (See Fig. \ref{Fig:PxtAll}). Also, a good agreement between $P_{XT}$ calculated by $UNIGE$ and $Nagoya$ can be found. However, we can notice, that results from $Catania$ for LVR-3050CS and LVR-6050CS devices show constantly higher $P_{XT}$, which is related to a difference in the data aqusition system and analysis procedure:
\begin{itemize}
\item $UNIGE$ and $Nagoya$ did measurements at high bandwidth (1GHz $UNIGE$ and 500 MHz $Nagoya$) with further offline analysis procedure which allows to eliminate the time window for pile-up effects down to 2 ns, while $Catania$ used bipolar shaper with 15 ns time constant;
\item $UNIGE$ and $Nagoya$ did offline correction for pile-up effect probability;
\end{itemize}

In the same time similar procedure was performed by $Catania$ only for Hexagonal device, while all other devices were measured using a 15 ns bipolar shaper at a temperature of $+2 \celsius$ where the chance coincidence of the thermal pulses is still significant.

\section{Conclusions}

In this paper we reported the initial work done in the framework of the SENSE project. In particular, related to work package 2: "$R \& D$ cooperation between academia and industry". Due to this agreement, in the first step, a comparison between $PDE$ and $P_{XT}$ measured by three partners (University of Geneva, Catania Observatory and Nagoya University) was performed, to crosscalibrate experimental setups as well as analysis procedures. The measurements were compared for three different SiPM devices with different active areas: from 9 $mm^ 2$ up to 93.6 $mm^2$ produced by Hamamatsu. 
It was found in average 7.8 $\%$ and 6.4 $\%$ relative difference in $PDE$ vs. $\Delta V$ and $PDE$ vs. $\lambda$ measuremets. By comparing the $P_{XT}$ measured by three partners, we can conclude that for precise $P_{XT}$ measurements the pile-up effect should be taken into account and properly eliminated.


\section{Acknowledgment}
This work was supported by the European Commission under Future and Emerging Technologies (FET) Open Coordination and Support Action (CSA), the Grant number 713171. We acknowledge support from JSPS/MEXT KAKENHI grant numbers 23244051, 25610040, 15H02086,16K13801 and 17H04838. The Italian participation to the work was supported by the Italian Ministry of Education, University, and Research(MIUR) with funds specifically assigned to the Italian National Institute of Astrophysics (INAF) for the Cherenkov Telescope Array (CTA), and by the Italian Ministry of Economic Development (MISE) within the “Astronomia Industriale” program. We acknowledge support from the Brazilian Funding Agency FAPESP (Grant 2013/10559-5) and from the South African Department of Science and Technology through Funding Agreement 0227/2014 for the South African Gamma-Ray Astronomy Programme. We gratefully acknowledge financial support from the agencies and organizations listed here: \url{http://www.ctaobservatory.org/consortium_acknowledgments}.

\end{document}